\documentclass[%
reprint,
superscriptaddress,
 amsmath,amssymb,
prl,
]{revtex4-2}
\usepackage{placeins}
\usepackage{graphicx}
\usepackage{dcolumn}
\usepackage{bm}
\usepackage{lipsum}
\usepackage{braket}
\usepackage[english]{babel}

\usepackage {xcolor}
\usepackage[T2A]{fontenc}
\usepackage{ulem}

\begin{document}

\preprint{APS/123-QED}

\title{
Super-resolved imaging based on spatiotemporal wavefront shaping 
}%

\author{Guillaume Noetinger}
\affiliation{ Institut Langevin, ESPCI Paris, Université PSL, CNRS, 75005 Paris, France }%
\author{Samuel Métais}
\affiliation{Aix Marseille Université, CNRS, Centrale Marseille, Institut Fresnel, Marseille, France}
\author{Geoffroy Lerosey}
\affiliation{ Greenerwave, 75002 Paris, France }%
\author{Mathias Fink}
\affiliation{ Institut Langevin, ESPCI Paris, Université PSL, CNRS, 75005 Paris, France }%
\author{Sébastien M. Popoff}
\affiliation{ Institut Langevin, ESPCI Paris, Université PSL, CNRS, 75005 Paris, France }%
\author{Fabrice Lemoult}
\affiliation{ Institut Langevin, ESPCI Paris, Université PSL, CNRS, 75005 Paris, France }
 \email{Corresponding author : fabrice.lemoult@espci.psl.eu}

\date{\today}

\begin{abstract}
A novel approach to improving the performances of confocal scanning imaging is proposed.
We experimentally demonstrate its feasibility using acoustic waves. 
It relies on a new way to encode spatial information using the temporal dimension. 
By moving an emitter, used to insonify an object, along a circular path, we create a temporally modulated wavefield.
Due to the cylindrical symmetry of the problem and its temporal periodicity, 
the spatiotemporal input field can be decomposed into harmonics corresponding to different spatial vortices, or topological charges.
Acquiring the back-reflected waves with receivers which are also rotating, 
multiple images of the same object with different Point Spread Functions (PSFs) are obtained. 
Not only is the resolution improved compared to a standard confocal configuration, 
but the accumulation of information also allows building images beating the diffraction limit. 
The topological robustness of the approach promises good performances in real life conditions. 
\end{abstract}

\maketitle

Imaging devices exploit waves to retrieve information about an object. 
Ultrasound imaging devices or optical microscopes are two widespread commercially-available technologies relying on different waves, 
both offering key insights for medical diagnosis and scientific research. 
While having different properties, those approaches rely on the same principles, and their resolution is limited by the same diffraction effects 
to a distance 
of the order of the wavelength. 
More specifically, 
in optical full-field microscopy, a sample is uniformly illuminated and the waves scattered off an object are collected with a microscope objective. The finite aperture of the optical system filters out waves corresponding to high scattering angles and thus the smaller details~\cite{AbbeLimit}. The image of a point, {\it i.e.} the Point-Spread Function (PSF) of the system, 
is an Airy spot whose first zero is located at a distance $1.22 \frac{\lambda}{2\:N\!A}$ from the focus 
(where $\lambda$ is the operating wavelength and $N\!A$ the numerical aperture). 
The Rayleigh criterion states that two point-like objects closer than this distance cannot be distinguished~\cite{rayleigh1879}.

Overcoming this limitation is the domain of {\it super-resolution} imaging and many techniques have already been proposed~\cite{SuperResReviewMicroscopy,LabelFreeSuperBook,liu2017plasmonics,lu2012hyperlenses,lemoult2010resonant,lemoult2012polychromatic,lemoult2011acoustic,SuperResReviewAcoustics,chaudhuri2001super,milanfar2017super}.
As any label-free imaging scheme can be decomposed in two main steps, 
{\it i.e.} acquiring data and processing the acquired data, 
those techniques can be divided in two different classes. 

The first strategy consists in shaping the illumination and/or the collection of waves. 
The simplest implementation is the well-known confocal microscope~\cite{Confocal,ConfocalReview},
where both the illumination and the detection correspond to a diffraction-limited volume. 
In addition to suppressing out-of-focus signals, 
it enhances the maximum transmitted spatial frequency by a factor of two~\cite{JMertz}. 
Nevertheless, such improvement is difficult to observe experimentally due to a poor gain for high spatial frequencies. 
The increase in lateral resolution is usually considered to be on the order of roughly 40\%. 
Other approaches were proposed, 
all relying on engineering the illumination or the collection scheme, 
such as diffractive tomography~\cite{Tomoreview,TomoWolf}, ptychography~\cite{PtychoFourier}, 
structured illumination~\cite{StructuredIll} or other PSF-engineering~\cite{PSFengineeringRoider,OptSuperOscMeta}. 
However, all these techniques remain ultimately limited in resolution by the diffraction.

The second strategy consists in developing algorithms for reconstructing the image.
Indeed, the Rayleigh criterion is somewhat an arbitrary rule. 
In particular, it assumes any spatial frequency outside of the transmitted bandwidth is definitely lost. 
Nonetheless, strong arguments exist to claim that the resolution can be limited only by the signal-to-noise ratio~\cite{FourierO,DiffractionResolutionLimit}. 
However, the presence of noise in real life experiments makes the problem ill-posed, 
forbidding a direct inversion of the mathematical equations. 
Various algorithms were developed to regularize the system by compensating for unknown or noisy information using strong priors~\cite{MicroscopyTech,UnlimitedResSLM,SuperResSparse,SuperResSingleShot}.

In this letter, we propose an approach combining wavefront shaping and mathematical deconvolution for improving the resolution. 
It consists in using dynamic wavefront shaping to rotate an illumination wavefront, 
thus using the temporal domain as an additional way to encode information. 
We show that it is equivalent to measuring the image of an object with multiple imaging systems with different orthogonal PSFs, 
corresponding to different harmonics of the received signals. 
The addition of information provided by those images allows increasing the effective signal to noise ratio and thus improving the resolution. 
We experimentally demonstrate this concept in acoustics in a confocal-like configuration 
by reconstructing the image of two small scatterers. 
Using a simple deconvolution process we show that two scatterers can be distinguished below the diffraction limit.

\begin{figure*}[t]
\includegraphics[width=\textwidth]{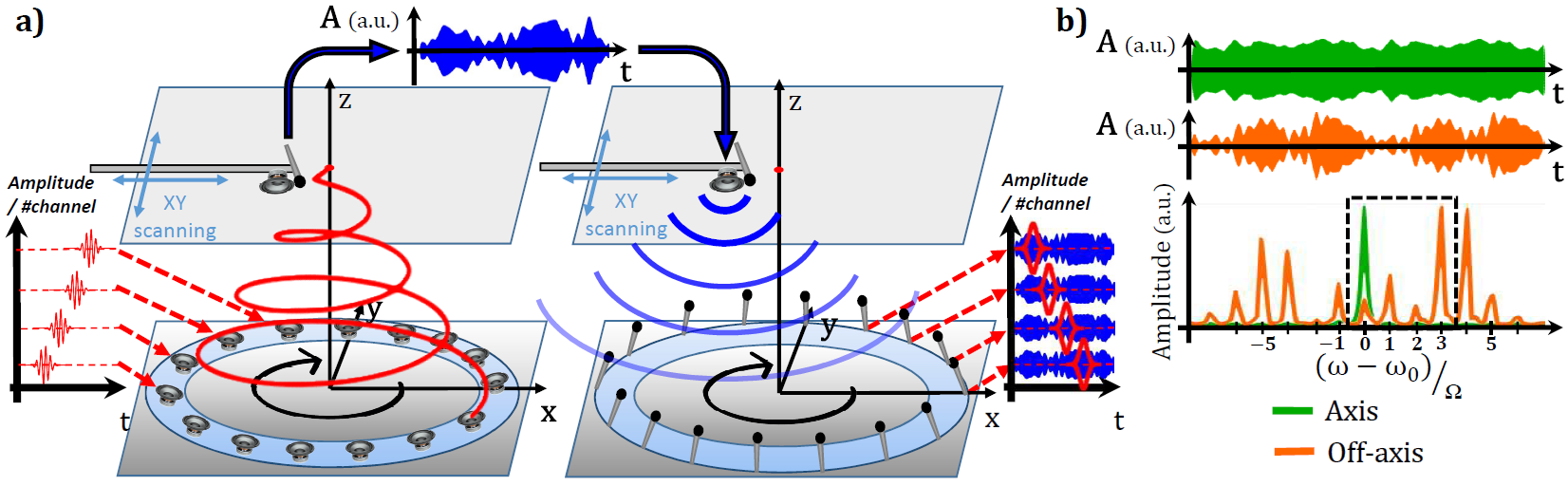}
\caption{\label{fig:Setup}\textbf{(a) Acoustic experimental setup. }
\textbf{Left} Loudspeakers, arranged in a circle, emit successively to mimic a single rotating loudspeaker. 
The pressure field is recorded in a \textit{focal plane} by a microphone on a translation stage. 
\textbf{Right} The previously recorded field is sent back by a loudspeaker next to the focal plane's microphone. 
This is equivalent to reflection by a point scatterer. 
The backscattered field is recorded by microphones arranged on a circle, 
which again mimic a rotating receiver by properly time-gating and summing all the signals. 
\textbf{ (b) Temporal signals and spectrums.}\label{fig:Spectrum} Backscattered signals for a scatterer on the rotation axis (top) and off-axis (middle). 
As shown by their spectrum (bottom), 
the on-axis signal is almost monochromatic while the off-axis one presents peaks at each harmonic $\omega_0+n\Omega$. \newline
In the next figure we examine the structure of the fields corresponding to $n \in [\![0;3]\!]$ in space. (black dotted box)}
\end{figure*}

The fundamental idea of the proposal relied on creating a singularity 
that would allow discriminating precisely one point from the rest of the object.
This is achieved using rotating wavefronts.
Point-like scatterers at different positions would perceive different time-modulations, 
depending on their distance to the rotation axis. 
Only a scatterer located exactly on the rotation axis backscatters an unmodulated wave. 
This modulation can also be understood using the Doppler effect: 
in the rotating frame, the object is rotating and thus Doppler shifts the waves. 
The further from the rotation axis, the more the frequency of the signal is shifted. 
Interestingly, as the rotation comes periodically to the same position, everything is actually encoded as a modulation on an initial monochromatic excitation.

For the sake of ease of access to the temporal field, 
we implement this concept with acoustic waves in the audible range, 
but the same framework could readily be transposed to other wave-based imaging systems, such as optical microscopes. 
In order to generate a rotating source, 
we place a set of 16 loudspeakers and microphones on a 55-cm-radius-circle. 
Each of them is connected to a multi-channel Antelope Orion32+ soundboard (see Fig.~\ref{fig:Setup}) 
controlled through a computer. 
The speakers emit sequentially a time-gated monochromatic signals 
in such a way that it is equivalent to having a
rotating source emitting a monochromatic signal at a frequency $\omega_0 = 1600~$Hz. 
(equivalent to a wavelength of $21.4$~cm). This has the advantage of not requiring any physical rotation.
The rotation angular frequency $\Omega = 4~$Hz is chosen so that the speed of the rotating emitter is low compared to the wave velocity. 
A microphone placed on a translation stage records the acoustic field in a plane (hereinafter called the \textit{focal plane}) located 1.7~meters above the speakers. 

To mimic a point-like scatterer re-emitting in all directions, 
a microphone and a loudspeaker are used to record the signal at a target position 
in the focal plane and then re-emit the recorded signal 
(Fig.~\ref{fig:Setup}\textbf{a} right). 
The microphones arranged on circle then measure the backscattered field. 
Finally, the recorded signals are time-gated and summed to mimic a rotating detector 
(see SI for experimental details).
We first show the transient back-scattered signals of a scatterer located 
on and outside the rotation axis in the focal plane (Fig.~\ref{fig:Spectrum}\textbf{b}).
As expected, 
the spectrum of the signal corresponding to the scatterer on the axis reveals to be almost monochromatic. On the contrary, the signal for the off-axis scatterer displays a periodic temporal modulation of its envelope. 
In the Fourier space, this corresponds to a frequency comb centered around $\omega_0$ 
with a spacing between peaks corresponding to the rotation frequency $\Omega$ (see bottom of Fig.~\ref{fig:Spectrum}b). 
Similarly, all the transient signals in this experiment display similar periodic modulation, and from now on  only the data corresponding to the different harmonics $\omega_0+n\Omega$, 
with $n \in \mathbb{Z}$, are displayed.

\begin{figure}[bt]
\includegraphics[width=0.5\textwidth]{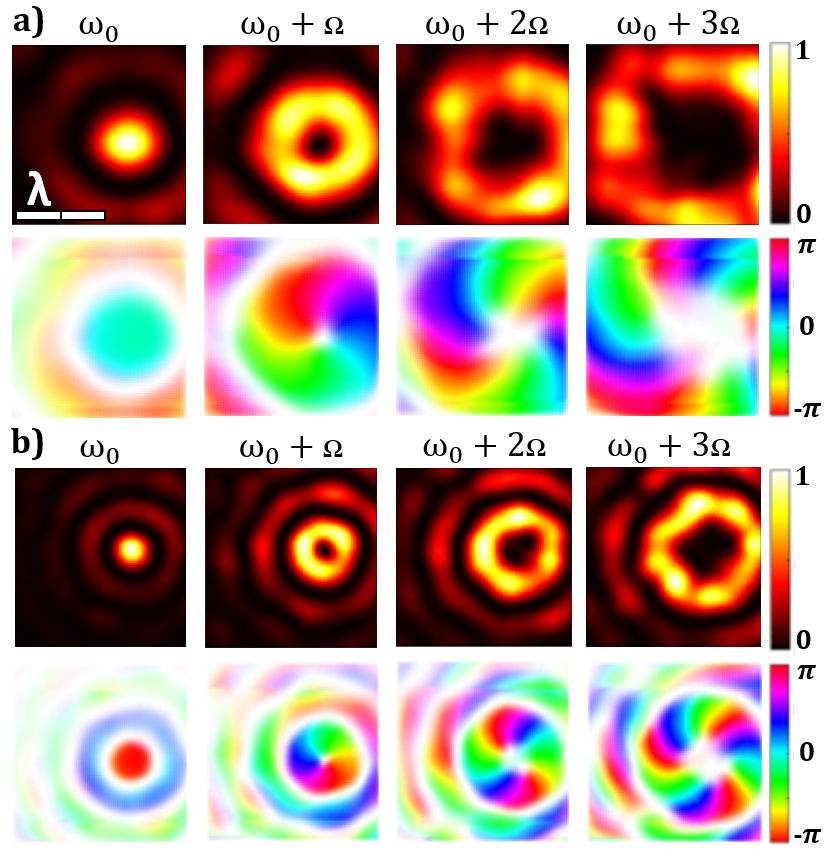} 
\caption{\label{fig:PSF_tot} \textbf{(a) Field in the focal plane} for the harmonics $\omega_0+n\Omega$ for $n\in [\![0;3]\!]$. \textbf{(b) Backscattered signals}\newline
For each dataset the first row corresponds to the intensity and the bottom row is the phase (weighted by the amplitude).}
\end{figure}

In order to investigate the imaging capabilities of our system, 
a full scan of the focal plane is performed. 
The field in the focal plane as well as the back-scattered signals are recorded. 
The spatio-temporal signals are then projected onto the different harmonics. 
The maps of the complex field in the focal plane corresponding to the harmonics in the range $[\![0;3]\!]$ are represented in the first part of Fig.~\ref{fig:PSF_tot}. 

The field at harmonic $n=0$, or equivalently at the initial carrier frequency $\omega_0$ (top first column of Fig.~\ref{fig:PSF_tot}), reveals a focal spot at the central position. Its width is approximately $1.3 \lambda$. 
It corresponds to the diffraction limited focal spot that would be obtained through a thin ring-aperture of $N\!A=0.31$ 
illuminated by a monochromatic wave at $\omega_0$.
Reciprocally, it corresponds to the $PSF$ of a full-field microscope with a thin ring-aperture, with a first zero at a distance $l_{\textrm{res}}=0.76 \frac{\lambda}{2\:N\!A}=28$~cm from the focal spot  (see SI).
The field at the harmonic $n$ is a vortex of topological charge $n$.
The field is concentrated on a ring whose radius increases with $|n|$. 
As expected, the central point who is on the rotation axis acts as a singularity. 
Similarly to the field at $\omega_0$, they are equivalent to the $PSF$ of a full-field acoustical microscope with a thin ring aperture on which a vortex plate would have been added.

The images corresponding to the first harmonics of the backscattered signals are also represented (bottom of Fig.~\ref{fig:PSF_tot}).
They strongly resemble to the maps in the focal plane, and notably they all exhibit similar vortices. 
However, there is a homothety relationship between the two set of images 
so that the rings corresponding to the maximum of intensity at each harmonic have a radius diminished by the same factor 2. 
This effect is a direct consequence of the dynamic configuration used in this experiment. It is similar to the doubling of the spatial frequency support obtained in structured illumination microscopy~\cite{StructuredIll} or diffractive tomography~\cite{Tomoreview}.

The next step consists in harnessing the new diversity of signals offered by the dynamic nature of the process 
to improve the imaging resolution below the diffraction limit. 
Our apparatus is equivalent to a confocal microscope in a monochromatic highly-coherent configuration,
except that the illumination pattern is modulated periodically in time.
As a result, the information of the object is encoded in the different harmonic frequencies. 
Consequently, we obtain as many images $\textrm{Im}_n(x,y)$ as the number of measured harmonic frequencies. 
Each of the images is also the convolution product between the object and the Point Spread Function $PSF_n(x,y)$,
which is specific to the harmonic $n$, 
and corresponds to a diffraction limited vortex of vorticity $n$. 
As two vortices with different topological charges are orthogonal, 
each channel carries a different information about the object.
Moreover, thanks to the reduced apparent wavelength previously observed in the backscattered $PSF$s, the resolution is increased. 

\begin{figure}[b]
\includegraphics[width=0.49\textwidth]{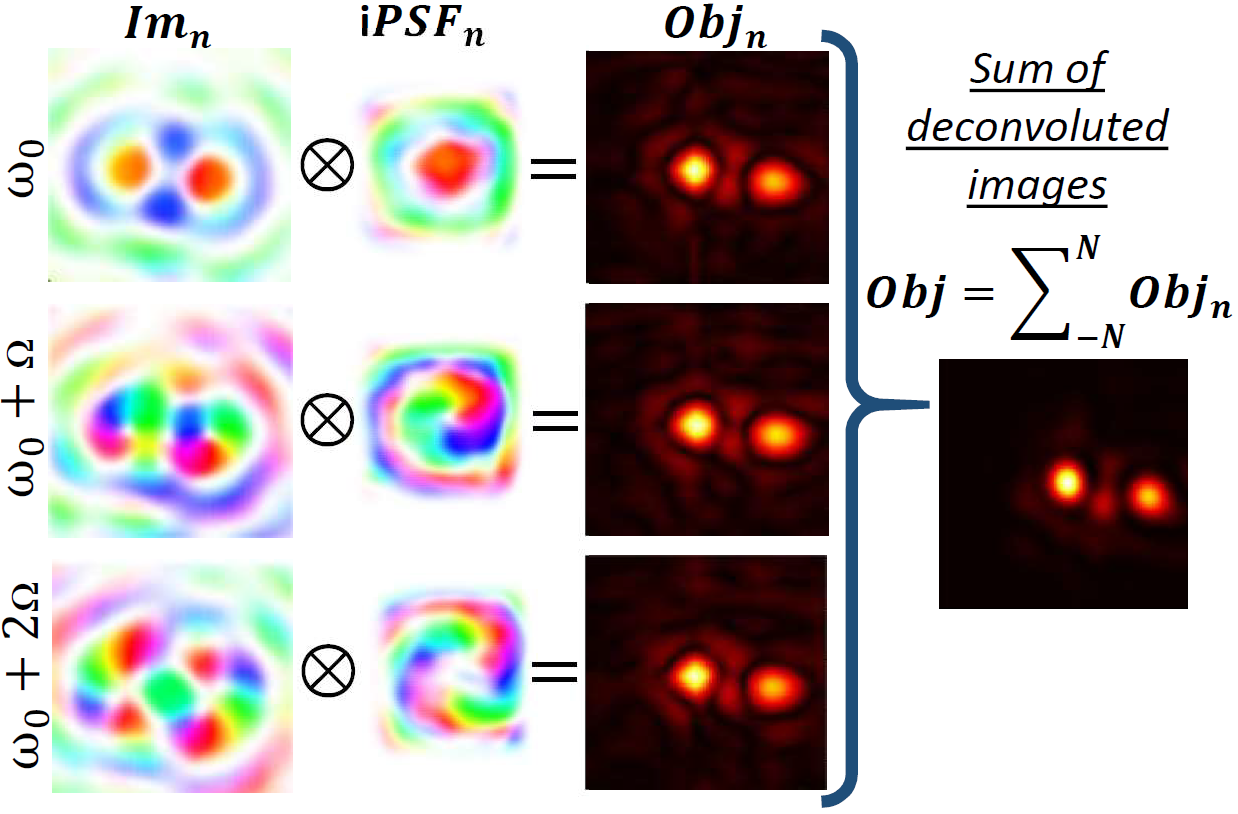}
\caption{\label{fig:Deconvolution} \textbf{Deconvolution procedure}. The images at each harmonic are convolved with a regularized inverse of the $PSF_n$
. All the independent reconstructions are summed to give a final sharp image.
}
\end{figure}
\begin{figure*}[t]
\includegraphics[width=\textwidth]{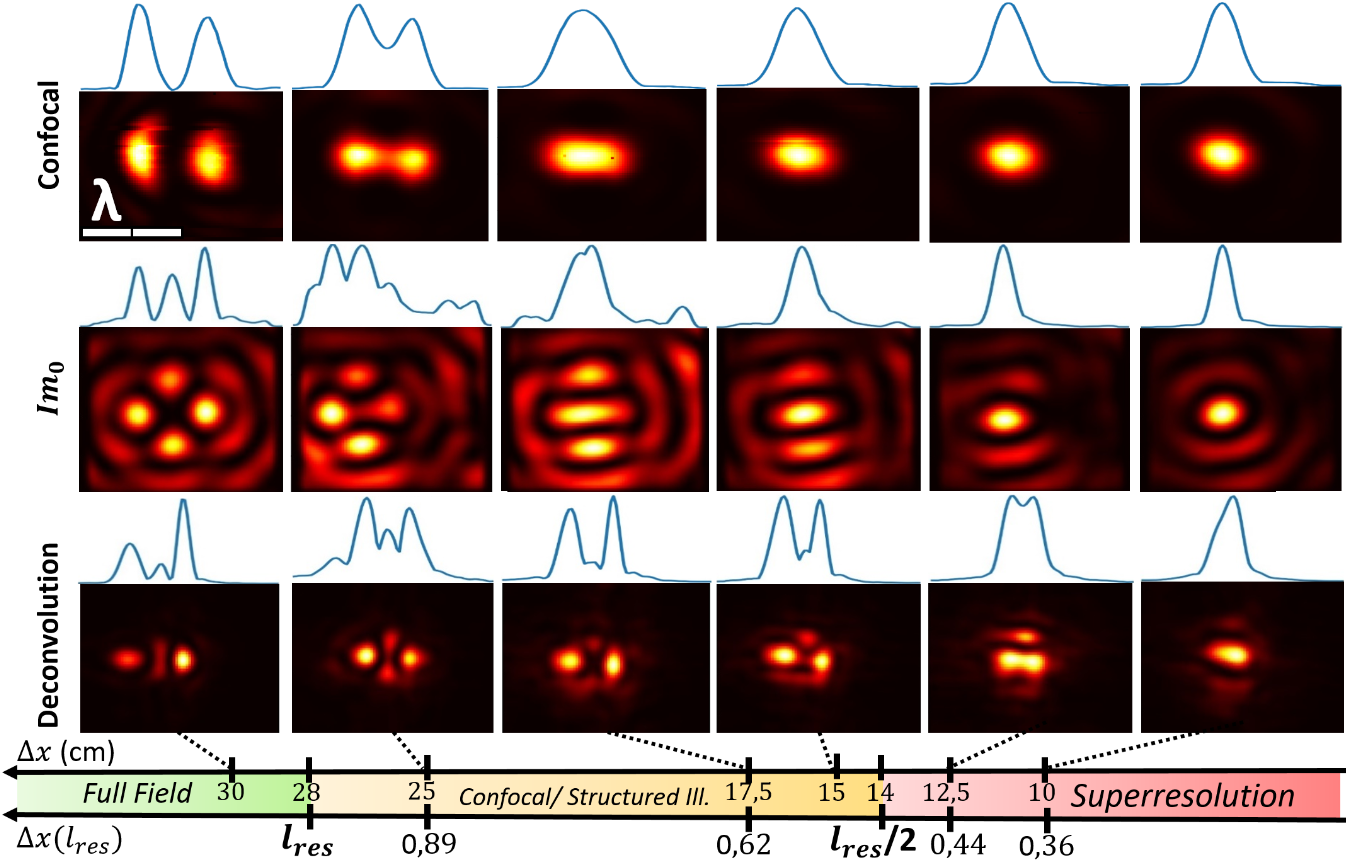}
\caption{ \label{fig:results} \textbf{Quantification of the resolution limit of the system} Two emitters in the focal plane are imaged in a confocal or dynamic way from full-field resolution to super-resolution regime. The field of view is approximately $4\lambda$ in X/Y. The images are shown in normalized intensity and the maximum profile on the separation axis (X) is shown above each image . \newline
Comparison between the confocal images (top row), the images at $\omega_0$ (middle row) and the deconvoluted images (bottom). 
Close to $\frac{l_{res}}{2}$, the confocal setup fails to distinguish two diffusers. 
In the dynamic images there are hints of the presence of the two diffusers in the signals: in the $\omega_0$ images shown they consists in the observed X/Y asymettry visible in the secondary lobes. 
With the deconvolution process the two point-like objects are made distinguishable further from this limit.}
\end{figure*} 
 
The images $\textrm{Im}_n$ corresponding to two point-scatterers separated by  $30$~cm are acquired.
The images of the first harmonics are displayed on the left of Fig.~\ref{fig:Deconvolution} as color coded images of their phase.
Each of them shows a vortex centered on each scatterer. 
The vortex has a topological charge that corresponds to the considered harmonic.

For each harmonic, a deconvolution is needed in order to obtain an image of the two scatterers. 
This consists in convolving the images by the inverse $iPSF_n$ of the point-spread function $PSF_n$ as depicted in Fig.~\ref{fig:Deconvolution}. 
Such an inverse is calculated from the experimental $PSF_n$ obtained with a single emitter in the spatial Fourier domain as it only involves dot products.
To be robust to the presence of noise, a Tikhonov regularization~\cite{TikhReg,TikhReg2,PopoffOpaqueTikh} is achieved. 
Each pseudo-inverse operator writes:
\begin{equation}
i\widehat{PSF}_n=(\widehat{PSF}_n^* \cdot \widehat{PSF}_n + \sigma)^{-1}\cdot \widehat{PSF}_n^* \label{InvTikh}
\end{equation}
\noindent where $\sigma$ corresponds to the noise-to-signal ratio, $\cdot^*$ denotes the complex conjugate 
and $\widehat{{\:.\:}}$ the 2D spatial Fourier transform. 
Note that in the real space, the inverse $iPSF_n$ exhibits a vorticity with a charge opposite to 
$PSF_n$.

By deconvolution, each harmonic provides its own intensity image of the two-scatterers. 
The last step consists in
summing all the images from harmonic $-4$ to $4$ and a final image is built (right of Fig.~\ref{fig:Deconvolution}). 
The signals add up and the average noise decreases leading to a cleaner final image of the two scatterers.

We then change the distance between the two scatterers to find when it becomes impossible to distinguish them. 
Fig.~\ref{fig:results} summarizes the results.
For the sake of generality, \textit{classical confocal images}, 
{\it ie.} where all the speakers emit simultaneously and all the microphones receive simultaneously,
are also performed. 
These images reveal that the two scatterers cannot be resolved when their distance is less than $22$~cm 
($\simeq \lambda = 0.79l_{\textrm{res}}$, corresponding to $30$\% of the diffraction limit).
The images at harmonic $n=0$ are represented in the central row.
They show smaller spatial variations but a high level of artifacts not allowing to distinguish any objects.
Eventually, when taking into accounts all the available harmonics and their inversion procedure, 
it is possible to beat the Rayleigh criterion and reach a resolution of $12.5~$cm = $0.58\lambda$. 
It corresponds to an  improvement of 70\% compared to the confocal configuration 
and 10\% compared to structured illumination 
($l_{\textrm{res}}/2$). 
Noise remains the limitation: the noise-to-signal ratio parameter $\sigma$, estimated at $0.003$ in those experiments, 
still limits the performance. 
This is a standard order of magnitude in microscopy~\cite{MicroscopyTech},
and any enhancement would immediately translate into an increase of the resolution capabilities. 
Moreover, further improvements can be made on the reconstruction using an iterative algorithm such as Jansson-Van Cittert~\cite{AlgoJansson} or a statistical one like Richardson-Lucy~\cite{ALgoRichardson} where our solution would be used as an initial guess
allowing one to envision enhanced performances of this approach.
In conclusion, we have proposed to exploit the time domain 
as a way to improve the resolution of an imaging device  
and demonstrated its applicability with an experiment based on audible acoustics. 
We showed that projecting a time-periodic pattern on a sample allows acquiring multiple images of the same sample with orthogonal $PSF$s. 
Using a rotating pattern, those $PSF$s correspond to vortices with different topological charges. 
The augmentation of the amount of information allows us to experimentally beat the diffraction limit, even in the presence of the inherent experimental noise. 
This new modality of imaging tackles the issue of diffraction in imaging systems in a novel way. 
In microscopy, it could offer an easy way to obtain label-free high or super resolution images.
This strategy could also be combined with non-linear phenomena such as fluorescence saturation involved in non-linear structured illumination~\cite{NLSIM}. 

This work is supported by LABEX WIFI (Laboratory of Excellence within the French Program “Investments for the Future”) under references ANR-10-LABX-24 and ANR-10-IDEX-0001-02 PSL*, and partially by  the Simons Foundation/Collaboration on Symmetry-Driven Extreme Wave Phenomena.


\providecommand{\noopsort}[1]{}\providecommand{\singleletter}[1]{#1}%

\end{document}